**Malignant field signature analysis in biopsy samples at diagnosis identifies lethal disease in patients with localized Gleason 6 and 7 prostate cancer.**

Gennadi V. Glinsky[1; 2]

[1] Institute of Engineering in Medicine

University of California, San Diego

9500 Gilman Dr. MC 0435

La Jolla, CA 92093-0435, USA

Correspondence: gglinskii@ucsd.edu

Web: http://iem.ucsd.edu/people/profiles/guennadi-v-glinskii.html

[2]Translational and Functional Genomics Laboratory, Genlighttechnology Corporation, La Jolla, CA 92037

Email: genlighttech@gmail.com

Web: www.genlighttechnology.com

**Abstract**

Overtreatment of early-stage low-risk prostate cancer (PC) patients represents a significant problem in disease management and has socio-economic implications. Development of genetic and molecular markers of clinically significant disease in patients diagnosed with low grade localized PC would have a major impact in disease management. A gene expression signature (GES) is reported for lethal PC in biopsy specimens obtained at the time of diagnosis from patients with Gleason 6 and Gleason 7 tumors in a Swedish watchful waiting cohort with up to 30 years follow-up. A 98-genes GES identified 89 and 100 percent of all death events 4 years after diagnosis in G7 and G6 patients, respectively; at 6 years follow-up, 83 and 100 percent of all deaths events were captured. Remarkably, the 98-genes GES appears to perform successfully in patients' stratification with as little as 2% of cancer cells in a specimen, strongly indicating that it captures a malignant field effect in prostates harboring cancer cells of different degrees of aggressiveness. In G6 and G7 tumors from PC patients of age 65 or younger, GES identified 86 percent of all death events during the entire follow-up period. In G6 and G7 tumors from PC patients of age 70 or younger, GES identified 90 percent of all death events 6 years after diagnosis. Classification performance of the reported in this study 98-genes GES of lethal PC appeared suitable to meet design and feasibility requirements of a prospective 4 to 6 years clinical trial, which is essential for regulatory approval of diagnostic and prognostic tests in clinical setting. Prospectively validated GES of lethal PC in biopsy specimens of G6 and G7 tumors will help physicians to identify, at the time of diagnosis, patients who should be considered for exclusion from active surveillance programs and who would most likely benefit from immediate curative interventions.

**Introduction**

In the United States, widespread implementation of the prostate-specific antigen (PSA) screening programs enabled diagnosis of more than 200,000 cases of prostate cancer each year (1). Clinically localized prostate cancer represents the vast majority of new cases (2). Therefore, one of the most significant benefits of the widespread use of PSA screening is that the prevalence of the late stage, advanced and high grade prostate cancer at diagnosis has declined dramatically and the vast majority of newly diagnosed prostate cancers are early stage and low grade tumors.

    The natural history of early stage clinically localized prostate cancer is considered favorable (2) and other types of cancer such as lung cancer are considered hundreds times as deadly. Despite this seemingly "indolent" nature, prostate cancer is the second leading cause of cancer-related deaths and accounts for 3.5% of all male deaths (3). Development of clear, consensus guidelines for physicians' decision-making process in clinical management of early stage localized prostate cancer is one of the most significant public healthcare problems. Inevitable and fast approaching demographic changes in the Western world underscore the critical economic and logistical needs for a rational, evidence-based approach to the clinical management of the early stage localized prostate cancer. A path to solutions to this problem is complicated by a multitude of competing positions attempting to emphasize the perceived shortcomings and benefits of different approaches and need to balance multiple variables such as public health care costs, individual patients' benefits, interests, socio-economic status, ethical and professional responsibilities of the medical personnel, and humanitarian considerations.

    Conclusive statistical evidence of the life-saving therapeutic benefits of radical prostatectomy versus watchful waiting in early prostate cancer have been documented in a randomized multicenter clinical trial: radical prostatectomy reduces disease-specific mortality, overall mortality, and the risks of metastasis and local progression (4-6). Immediate curative interventions are the predominant

therapy choice and 168,000 prostatectomies are performed each year to treat prostate cancer (7). It seems reasonable to conclude, that early detection of prostate cancer facilitated by PSA screening and aggressive use of radical prostatectomy for treatment of early prostate cancer have contributed to a significant extent to the reported 98-100% five-year survival rates since 1998 in the United States (SEER 13 areas statistics).

However, there is a lack of consensus regarding the benefits of a population-scale PSA screening and a controversy about the potential for overdiagnosis and overtreatment of clinically insignificant disease that would not likely to become life-threatening in a man's lifetime (8). Further socio-economic arguments in support of significant overdiagnosis and overtreatment have been presented in studies indicating that prevention of one prostate cancer death would require active treatment of 48 men for nine years or 12 men for 14 years (9, 10). Outcome studies from contemporary population-based cohorts reported cumulative 10-year prostate cancer-specific mortality in patients with low-risk disease 2.4% and 0.7% in the surveillance group and curative intent groups, respectively (11), which indicates that the surveillance may be a suitable treatment option for majority of patients with low-risk prostate cancer. Clinical evidence that active surveillance may be a safe, perhaps preferred option for older men diagnosed with a very low-grade or small-volume form of prostate cancer were published recently by Carter and colleagues (12). Therefore, active surveillance with curative intent for low-risk prostate cancer is under active consideration as a potentially safe alternative to immediate curative intervention with the expectations that it may reduce overtreatment and therapy-associated adverse events. It certainly would reduce the escalating economic burden of cost of prostate cancer treatment. The major limitation of these studies is a short follow-up time [for example, in the John Hopkins study (12), the total cohort has a median follow-up of 2.7 years (range 0.01 to 15)] which requires the use of biochemical recurrence or other "proxy" end-points for disease-specific mortality. This limitation is particularly relevant for early prostate cancer because the overall survival benefits of radical prostatectomy versus watchful waiting are not statistically apparent until 10 years follow-up (4-6) due to the fact that a majority of death events in the watchful waiting cohorts of early prostate cancer occurs at or after 10 years follow-up (4-6; this study). Furthermore, significantly longer follow-up data are required because most patients currently diagnosed with localized prostate cancer are aged 60–70 years and have a life expectancy of more than 15 years (11). Most importantly, there are no genetic or molecular methods prospectively defining low-risk or indolent prostate cancer at diagnosis with sufficient specificity and selectivity to ensure the safety of patients and allow physicians to make informed, ethical, evidence-based disease management decision of not treating prostate cancer. Given the natural history of early prostate cancer and long-term survival data from watchful waiting cohorts, conclusive prospective validation of laboratory methods defining low-risk indolent disease in Gleason 6 and 7 patients would require at least 10 years. Based on the analysis of the long-term survival data of prostate cancer patients from watchful waiting cohorts with up to 30 years follow-up, we reasoned that more feasible and clinically-relevant approach would be an attempt to identify genetic markers of lethal prostate cancer in patients with Gleason 6 and 7 tumors which would capture a vast majority of all cancer-related death events 4-6 years after diagnosis. Here we report identification of gene expression signatures (GES) of lethal prostate cancer in biopsy specimens obtained at the time of diagnosis from patients with Gleason 6 and 7 tumors in a Swedish watchful waiting cohort with up to 30 years follow-up. In retrospective analysis, best-performing GES of lethal prostate cancer identify 89% and 100% of all death events 4 years after diagnosis in Gleason 7 and Gleason 6 patients, respectively. GES appear to perform successfully in patients' stratification with as little as 2% of cancer cells in a specimen. In Gleason 6 and 7 prostate cancer patients of age 65 or younger, GES identifies 86% of all death events during the follow-up. In Gleason 6 and 7 prostate cancer patients of age 70 or younger, GES identifies 90% of all death events 6 years after diagnosis. Reported in this study GES of lethal prostate cancer in biopsy specimens of Gleason 6 and 7 tumors should help practicing physicians to identify at the time of diagnosis prostate cancer patients who should be considered for exclusion from the active surveillance programs and who would most likely benefit from immediate curative interventions.

**Materials and Methods**

**Patients**

This study is based on prostate cancer patients from the population-based Swedish Watchful Waiting cohort of men with localized prostate cancer (4-6, 13). Distinguishing feature of this cohort is that it represents patients diagnosed with symptomatic early prostate cancer at the time when no PSA screening programs were in place: these men had symptoms of benign prostatic hyperplasia (lower urinary tract symptoms) and were subsequently diagnosed with prostate cancer. All men in this study were determined at the time of diagnosis to have clinical stage T1 and T2, Mx, and N0, according to the 2002 American Joint Commission Committee TNM staging system (4-6, 13). The prospective follow-up time in this cohort is now up to 30 years and the study cohort was followed for cancer-specific and all cause mortality until March 1, 2006 (10). Deaths were classified as cancer-specific when prostate cancer was the primary cause of death as determined through a complete review of medical records by a study end-point committee (4-6, 13). Importantly, that in addition to the histopathological examination at the time of diagnosis, slides and corresponding paraffin-embedded formalin-fixed blocks were subsequently retrieved and re-reviewed to confirm cancer status and to assess Gleason scores using review, examination, and grading procedures blinded with regard to disease outcome (13).

**Gene expression analysis, evaluation, and selection of gene expression signatures**

Gene expression signatures (GES) were developed based on a publicly available microarray analysis of a Swedish Watchful Waiting cohort with up to 30 years of clinical follow up using a novel method for gene expression profiling [cDNA-mediated annealing, selection, ligation, and extension (DASL) method] which enabled the use of formalin-fixed paraffin-embedded transurethral resection of prostate (TURP) samples taken at the time of the initial diagnosis. Details of the experimental procedure can be found in a recent publication (13) and in Gene Expression Omnibus (GEO: <http://www.ncbi.nlm.nih.gov/geo/> ) with platform accession number: GPL5474. Full data set and associated clinical information is available at GEO with accession number: GSE16560.

Feature selection was performed without assessment of differential gene expression between deceased and surviving patients. All 6144 genes were evaluated for association with clinical and pathological variables (except survival status) using correlation analysis. Different thresholds on the p-values (0.05; 0.01; 0.001) were used for selection of gene sets with common patterns of association and concordance analysis was performed using expression profiling data of snpRNA-driven cell line-based models of prostate cancer predisposition (14, 15) to identify concordant and discordant gene expression signatures in cell lines and clinical samples (16-19). GES were built based on selection of co-regulated transcripts in various experimental conditions and clinically-relevant models, including prostate cancer predisposition and longevity models (14-19). Underlying concept at this stage of the analysis was to identify GES with concordant expression profiles across multiple data sets (16-19). Cox regression analysis was carried out to identify statistically significant candidate GES associated with patients' survival status. Cut-off threshold of p-values was set based on the p-value of the best-performing clinico-pathological parameter (Gleason score) in univariate Cox regression analysis (p = 0.0113). Genes from statistically significant GES were split, combined, and permutated using random iteration process to find novel statistically significant combinations based on univariate Cox regression analysis. GES scores were derived directly from measurements of expression values of each gene by calculating a single numerical value for each patient. GES scores represent the difference between sums of expression values of genes with common co-regulation profiles which is defined by up-regulation and/or positive correlation values versus down-regulation and/or negative correlation values. GES with p values < 0.01 were selected for further evaluation using multivariate Cox regression analysis of classification models which include GES and clinico-pathological co-variats (age and Gleason score). Cut-off threshold of p-values for candidate GES selection was set based on the p-value of the best-performing clinico-pathological model (age and Gleason score) in multivariate Cox regression analysis (p = 0.0052). Candidate GES that outperformed clinico-pathological models in multivariate Cox regression analysis were selected for further consideration using a split-sample validation procedure for classification threshold selection and GES classification performance evaluation as previously described (16-19).

Gene expression-based classification models were designed and evaluated through a split-sample validation procedure which enables the unbiased estimation of the performance of a classifier since the evaluation is performed on an independent data set (20). Specifically, the entire data set of 281 patients was split into training and test sets (141 and 140 patients, respectively), with approximately equal proportion of men with lethal and indolent prostate cancer and statistically undistinguishable clinical and pathological variables, e.g., age and time of diagnosis, follow up time, Gleason scores, percent of cancer cells in specimens (Table 1). The training set of 141 samples was utilized to identify and select the best classifier, whose performance was evaluated on the test set of 140 samples without any further adjustments to the threshold selection and classification protocols using Kaplan-Meyer survival analysis essentially as previously described (16-19). Best-performing GES classifiers were further evaluated in various clinically-relevant patients' sub-groups, including only Gleason 6 patients (n = 83), only Gleason 7 patients (n = 117), Gleason 6 and 7 patients (n = 200), with further sub-division of patients in additional validation screens based on age at diagnosis (age 65 and younger; age 70 and younger) and percent of cancer cells in the samples (2%; 5% or less; 10% or less; 20% or less; 40% or less; and 50% or more). In all these secondary validation screens no further adjustments to the threshold selection and classification protocols were made. 98 genes classifier that remains statistically significant in all these validation screens is reported in this paper.

Statistical significance of the Pearson correlation coefficients for individual test samples, clinical variables, and the appropriate reference standard were determined using GraphPad Prism version 4.00 software. We calculated the significance of the differences in the numbers of death events and surviving patients between the groups using two-sided Fisher's exact test and the significance of the overlap between the lists of differentially-regulated genes using the hypergeometric distribution test (21).

**Results and discussion**

Clinical characteristics of the training and test sets are provided in Table 1, and further details for the entire Swedish Watchful Waiting cohort are available in a recent publication (13) and in Gene Expression Omnibus (GEO: http://www.ncbi.nlm.nih.gov/geo/ ) with accession number GSE16560. All of the 281 patients in the Swedish cohort had clinical symptoms and were diagnosed from TURP or adenoma enucleation samples and thus were staged depending on the proportion of the tissue that was cancerous either T1a or T1b (13). Analysis of survival data in the entire cohort of 281 patients indicates that prostate cancer patients with different Gleason scores have markedly distinct timelines of death events during the extended up to 30 years follow-up (Figure 1). Most striking indicator is that only 6% of untreated Gleason 6 prostate cancer patients died at 5 years; 14% died between 5 to 10 years; and a majority of deaths (~ 35%) occurs 10 – 23 years after diagnosis. This analysis suggests that a majority of all death events (> 60%) in untreated Gleason 6 prostate cancer patients is occurring more than 10 years after diagnosis and during the sufficiently long follow-up period more than 50% of these patients will die (Figure 1). Long-term survival timelines for untreated Gleason 7 prostate cancer patients with symptomatic prostate cancer appear even more alarming: 27% died at 5 years follow-up; 22% of deaths occurred between 5 to 10 years; and > 70% died during the entire follow-up period (Figure 1).

Collectively, the analysis of timelines of death events in a watchful waiting cohort indicates that a majority of patients with symptomatic Gleason 6 and 7 prostate cancers will eventually develop clinically significant disease during sufficiently long follow-up period which further underscore the critical need to reliably define lethal prostate cancer at diagnosis. We applied the univariate Cox regression analysis to the entire cohort of 281 patients to identify several GES with the p value < 0.01 which appear to perform better than the best clinico-pathological co-variate, Gleason score (p = 0.0113; Supplemental Table S1). Most of these GES outperformed the clinico-pathological classification model in multivariate Cox regression analysis as well (Supplemental Table S2).

Separating the cohort of 281 patients into training and test cohorts and using the Kaplan-Meier survival analysis, we identified 98 genes GES that manifest the highly significant classification performance in the training set, retained highly consistent classification performance in the test set, and remained a highly significant classifier in the pooled cohort (Figure 1). It is important to note that in all secondary validation screens following the training set analysis no further adjustments to the threshold selection and classification protocols were made.

Notably, prostate cancer patients with identical Gleason scores (e.g., Gleason 6 patients and Gleason 7 patients) which were segregated into lethal and moderate disease sub-groups based on 98 genes GES classification had highly significant differences in the survival rates (Figure 1). These data suggest that 98 genes GES may be useful in identifying lethal disease in patients diagnosed with low grade localized prostate cancer. To test this hypothesis, we performed Kaplan-Meier survival analysis based on 98 genes GES classification in the cohort of 200 patients with Gleason 6 and 7 prostate cancer (Figure 2). We found that 98 genes GES is a highly significant classifier of Gleason 6 and 7 prostate cancer patients into sub-groups with lethal and moderate disease (Figure 2). 98 genes GES of lethal prostate cancer performs as a highly significant after segregation of patients into separate Gleason 6 and Gleason 7 sub-groups: 89% and 100% of all death events were identified 4 years after diagnosis in Gleason 7 and Gleason 6 patients, respectively; at 6 years follow-up, 83% and 100% of all deaths events were captured in Gleason 7 and 6 patients, respectively (Figure 2).

Age at diagnosis is considered among very important clinical determinants guiding the decision making process in clinical management of prostate cancer. This is particularly important for relatively younger patients because patients diagnosed with prostate cancer at age < 65 years are more likely to benefit from the immediate curative therapies (6). We therefore attempted to determine whether 98 genes GES will identify lethal disease in prostate cancer patients of differing ages. Remarkably, Kaplan-Meier survival analysis has determined that 98 genes GES performed very efficiently in stratification of prostate cancer patients of 65 years or younger (Figure 3): in Gleason 6 and 7 prostate cancer patients of age 65 or younger, GES identifies 86% of all death events during the follow-up. In Gleason 6 and 7 prostate cancer patients of age 70 or younger, GES identifies 90% of all death events 6 years after diagnosis (Figure 3).

Proportion of cancer cells in biopsy samples is highly variable and these variations may have significant impact on performance of gene expression-based classifiers. In biopsy samples from the population-based Swedish Watchful Waiting cohort the reported percent of cancer cells in a sample varied dramatically from 2% to 90%. We therefore set out to determine whether the number of cancer cells in biopsy samples would have an impact on classification performance of the 98 genes GES of lethal prostate cancer. We applied the 98 genes GES classifier to prostate cancer patients which were segregated into distinct sub-groups based on the percent of cancer cells in a biopsy sample. Kaplan-Meier survival analysis demonstrates that 98 genes GES performs successfully in patients' stratification regardless of the number of cancer cells in biopsy samples (Figures 4 & 5). Remarkably 98 genes GES appear to identify lethal disease in Gleason 6 and 7 prostate cancer patients with as little as 2% of cancer cells in a biopsy specimen (Figure 5). The conclusions reached based on the Kaplan-Meier survival analyses were confirmed using the Receiver Operating Characteristic (ROC) area under the curve analysis of the patients' classification based on the 98-genes signature score in training (n = 141) and test (n = 140) groups (A) and different clinically-relevant sub-groups (B - D) of patients (Figure 6; Tables 2 & 3). Collectively, the results of the present analyses strongly indicate that the 98-genes GES captures a malignant field effect in the human prostates harboring cancer cells with markedly different clinical aggressiveness.

    The most recent beta release of web-based tools, the UCSC Xena (http://xena.ucsc.edu/ ), provides powerful resources to explore, analyze, and visualize the comprehensive functional cancer genomics datasets of thousands annotated clinical samples of the Cancer Genome Anatomy Project (TCGA) (https://genomecancer.soe.ucsc.edu/proj/site/xena/datapages/ ). The classification performance of the 98-genes GES was further validated using TCGA Prostate Cancer cohort of 550 clinical samples with known therapy outcomes after the initial treatment (Table 4). Importantly, tumors tissues of the TCGA cohort comprise the prostatectomy samples which were analyzed using the state of the art Illumina Next Generation Sequencing technology.

    Decision making process in clinical management of low-risk localized prostate cancer is likely to affect life and death of thousands of patients. The problem is confounded by the fact that statistically significant survival benefits of curative therapy are evident only 10 years after diagnosis of the early-stage prostate cancer. Therefore, any genetic or molecular tests designed to aid physicians and patients in this process would require the regulatory approval following the successful prospective clinical trial. Classification performance of the reported in this study 98 genes GES of lethal prostate cancer appears highly suitable to meet design and feasibility requirements of the prospective 4 to 6 years clinical trial. Prospectively validated GES of lethal prostate cancer in biopsy

specimens of Gleason 6 and 7 tumors will help practicing physicians to identify at the time of diagnosis individual patients who should be considered for exclusion from the active surveillance programs and who would most likely benefit from the immediate curative interventions.

**Supplemental Information**

Supplemental Tables 1-3 are presented at the end of this manuscript. Additional supplemental information is available upon request.

**Competing Interests**

No competing interests and potential conflicts of interest were disclosed.

**Author Contributions**

This is a single author contribution. All elements of this work, including the conception of ideas, formulation, and development of concepts, execution of experiments, analysis of data, and writing of the paper, were performed by the author.

**Acknowledgements**

This work was made possible by the open public access policies of major grant funding agencies and international genomic databases and the willingness of many investigators worldwide to share their primary research data. I would like to thank you many colleagues for their valuable critical contributions during the preparation of this manuscript.

**Table 1.** Clinical characteristics of prostate cancer patients in the training and test sets

| Characteristic | Training set (n = 141) | Test set (n = 140) |
|---|---|---|
| Years of diagnosis, range (years) | 1977-1998 | 1977-1998 |
| Years of diagnosis, years (mean +/-SD) | 1991 +/- 4.1 | 1991 +/- 4.0 |
| Age at diagnosis, range (years) | 51-91 | 55-91 |
| Age at diagnosis, years (mean +/-SD) | 74.5 +/- 7.5 | 73.5 +/- 7.0 |
| Follow-up time, range (months) | 6-274 | 7-259 |
| Follow-up time, months (mean +/-SD) | 102.3 +/- 57.2 | 101.9 +/- 55.7 |
| Percent of cancer in samples, range (%) | 2% - 90% | 2% - 90% |
| Percent of cancer in samples, % (mean +/-SD) | 22.9 +/- 22.7 | 24.0 +/- 25.5 |
| Gleason scores, number (%) | | |
| Gleason 6 | 42 (29.8) | 41 (29.3) |
| Gleason 7 | 62 (44) | 55 (39.3) |
| Gleason 8-10 | 37 (26.2) | 44 (31.4) |
| Clinical outcomes, number (%) | | |
| Deceased | 105 (74.5) | 101 (72.1) |
| Alive | 36 (25.5) | 39 (27.9) |

**Table 2.** ROC area under the curve analysis of training and test data sets.

| Data sets and survival time | 10 yrs | 7 yrs | 6 yrs | 5 yrs | 4 yrs |
|---|---|---|---|---|---|
| Training set (n = 141) | 0.85 | 0.854 | 0.814 | 0.788 | 0.794 |
| Test set (n = 140) | 0.826 | 0.801 | 0.786 | 0.758 | 0.759 |

**Table 3.** Percent of all death events at different follow-up time in lethal prostate cancer groups of training and test data sets.

| Data sets and survival time | 10 yrs | 7 yrs | 6 yrs | 5 yrs | 4 yrs |
|---|---|---|---|---|---|
| Training set (n = 141) | 75% | 83% | 82% | 84% | 84% |
| Test set (n = 140) | 83% | 88% | 87% | 84% | 84% |

**Table 4.** Classification performance of the 98-genes GES in the TCGA cohort of 550 prostate cancer patients with known therapy outcomes after the initial treatment.

| Categories | Therapy outcomes after the initial treatment (number of patients with adverse events) | | |
|---|---|---|---|
| Patients' sub-group/ Adverse events | Relapse | Biochemical recurrence | New tumors |
| Poor prognosis (n = 275) | 33 | 44 | 60 |
| Good prognosis (n = 275) | 10 | 18 | 20 |
| Patients' sub-group/ Adverse events | Therapy outcomes after the initial treatment (percent of patients with adverse events) | | |
| Poor prognosis (top 50% scores) | 12.00 | 16.00 | 21.82 |
| Good prognosis (bottom 50% scores) | 3.64 | 6.55 | 7.27 |
| P value* | 0.0004 | 0.0006 | <0.0001 |

Legend: *P values were estimated using 2-talied Fisher's exact test. TCGA, the Cancer Genome Anatomy Project. At the date of the analyses, the median follow-up time in the prostate cancer TCGA cohort was 2.1 years.

**Figure legends**

**Figure 1.** Natural history of prostate cancer progression in patients' population from a Swedish watchful waiting cohort with up to 30 years follow-up (A) and classification performance of the 98 genes signature of lethal disease in prostate cancer patients (B-E). A, cancer-specific survival data in the entire watchful waiting cohort are presented to illustrate markedly distinct survival timelines of non-treated prostate cancer patients diagnosed with different Gleason scores prostate cancer. Kaplan-Meier survival analysis of the classification performance of the 98 genes GES in the training set (B), test set (C), and pooled cohort of 281 patients (D, E). Classification threshold 98 genes GES score of 270.43 units was chosen using the training set of 141 prostate cancer patients and consistently applied in all subsequent validation screens using the Kaplan-Meier survival analysis to stratify the patients into lethal disease sub-groups (score >= 270.43) and moderate/aggressive disease sub-group (score < 270.43). Percent value indicates the proportion of patients in the lethal disease sub-group. P values indicate the significance of the differences in the numbers of death events and surviving patients between the groups which was determined using two-sided Fisher's exact test.

**Figure 2.** Gene expression signature-based identification of lethal disease in Gleason 6 and 7 prostate cancer patients. Kaplan-Meier survival analysis of the classification performance of the 98 genes GES in 200 Gleason 6 and 7 prostate cancer patients (A), 83 Gleason 6 patients (B), and 117 Gleason 7 patients (C). Classification threshold 98 genes GES score of 270.43 units was chosen using the training set of 141 prostate cancer patients and consistently applied in all subsequent validation screens using the Kaplan-Meier survival analysis to stratify the patients into lethal disease sub-groups (score >= 270.43) and moderate/aggressive disease sub-group (score < 270.43). Percent values indicate the proportion of patients in the lethal disease sub-group. P values indicate the significance of the differences in the numbers of death events and surviving patients between the groups which was determined using two-sided Fisher's exact test.

**Figure 3.** Gene expression signature-based identification of lethal disease in prostate cancer patients with different age at diagnosis. Kaplan-Meier survival analysis of the classification performance of the 98 genes GES in 34 prostate cancer patients of age 65 or younger (A), 64 prostate cancer patients of age 70 or younger (B). Bottom figures in both A and B panels show the results of Kaplan-Meier survival analysis for Gleason 6 and 7 patients only of corresponding age groups. Classification threshold 98 genes GES score of 270.43 units was chosen using the training set of 141 prostate cancer patients and consistently applied in all subsequent validation screens using the Kaplan-Meier survival analysis to stratify the patients into lethal disease sub-groups (score >= 270.43) and moderate/aggressive disease sub-group (score < 270.43). Percent values indicate the proportion of patients in the lethal disease sub-group. P values indicate the significance of the differences in the numbers of death events and surviving patients between the groups which was determined using two-sided Fisher's exact test.

**Figure 4.** Gene expression signature-based identification of lethal disease in prostate cancer patients with distinct numbers of cancer cells in biopsy samples. Kaplan-Meier survival analysis of the classification performance of the 98 genes GES in 59 prostate cancer patients having 2% cancer cells in biopsy samples (A, top), 91 patients having 5% or less cancer cells in biopsy samples (A, bottom), 135 patients having 10% or less cancer cells in biopsy samples (B, top), 180 patients having 20% or less cancer cells in biopsy samples (B, bottom; and C, top), 220 patients having 40% or less cancer cells in biopsy samples (C, bottom). Classification threshold 98 genes GES score of 270.43 units was chosen using the training set of 141 prostate cancer patients and consistently applied in all subsequent validation screens using the Kaplan-Meier survival analysis to stratify the patients into lethal disease sub-groups (score >= 270.43) and moderate/aggressive disease sub-group (score < 270.43). Percent values indicate the proportion of patients in the lethal disease sub-group. P values indicate the significance of the differences in the numbers of death events and surviving patients between the groups which was determined using two-sided Fisher's exact test.

**Figure 5.** Gene expression signature-based identification of lethal disease in Gleason 6 and 7 prostate cancer patients with distinct numbers of cancer cells in biopsy samples. Kaplan-Meier survival analysis of the classification performance of the 98 genes GES in 52 prostate cancer patients having 2% cancer cells in biopsy samples (A, top), 76 patients having 5% or less cancer cells in biopsy samples (A, bottom), 109 patients having 10% or less cancer cells in biopsy samples (B, top), 140 patients having 20% or less cancer cells in biopsy samples (B, bottom; and C, top), 167 patients having 40% or less cancer cells in biopsy samples (C, bottom). Classification threshold 98 genes GES score of 270.43 units was chosen using the training set of 141 prostate cancer patients and consistently applied in all subsequent validation screens using the Kaplan-Meier survival analysis to stratify the patients into lethal disease sub-groups (score >= 270.43) and moderate/aggressive disease sub-group (score < 270.43). Percent values indicate the proportion of patients in the lethal disease sub-group. P values indicate the significance of the differences in the numbers of death events and surviving patients between the groups which was determined using two-sided Fisher's exact test.

**Figure 6.** Receiver Operating Characteristic (ROC) area under the curve analysis of the patients' classification based on the 98-genes signature score in training (n = 141) and test (n = 140) groups (A) and different clinically-relevant sub-groups (B - D) of patients.

**Figure 1.**

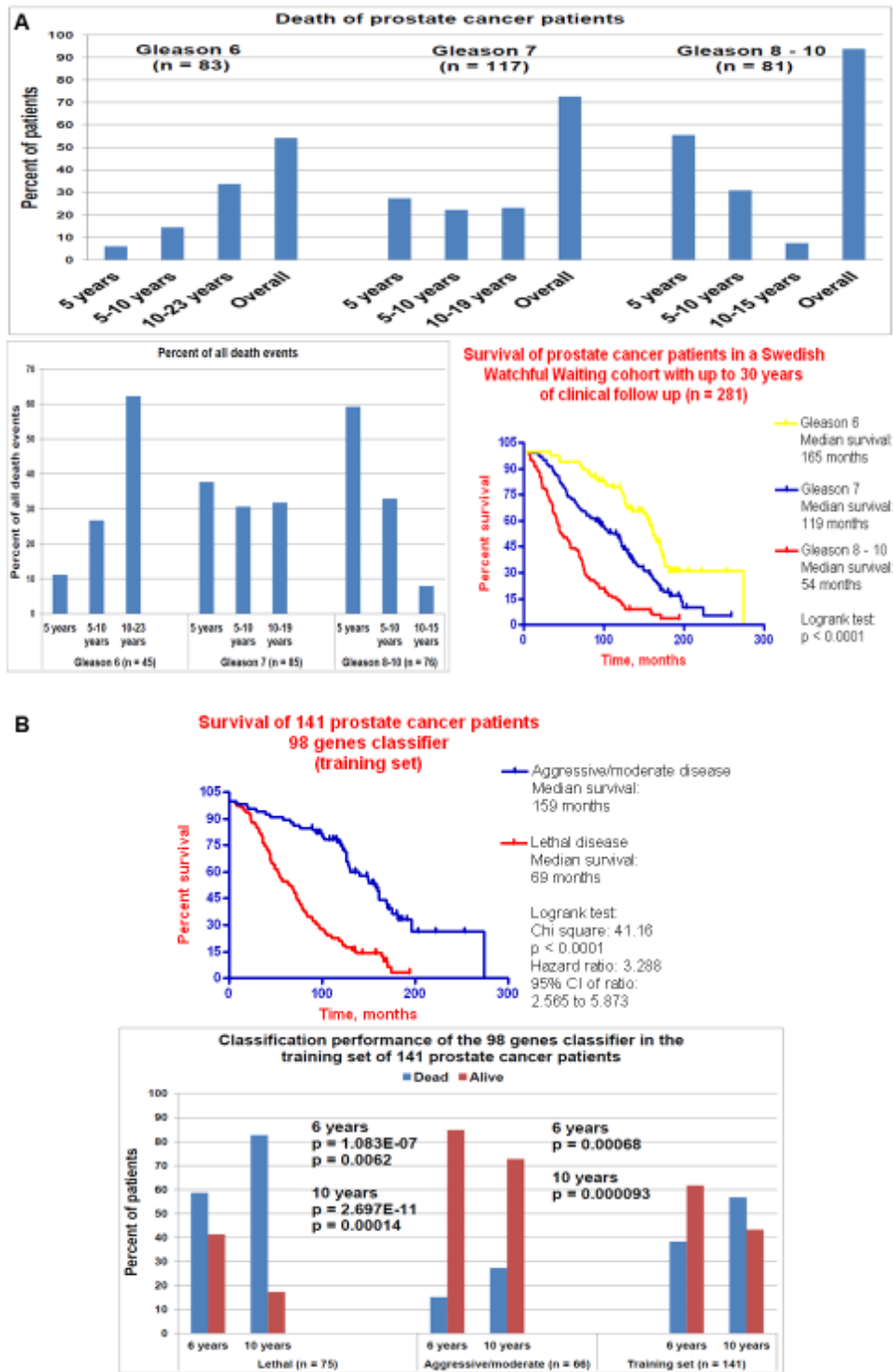

C

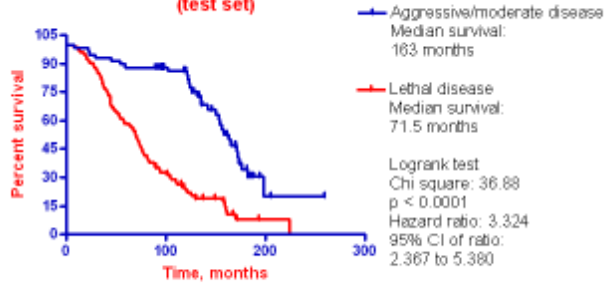

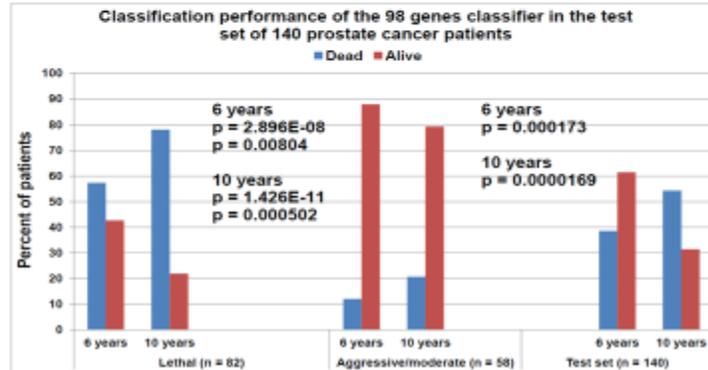

D

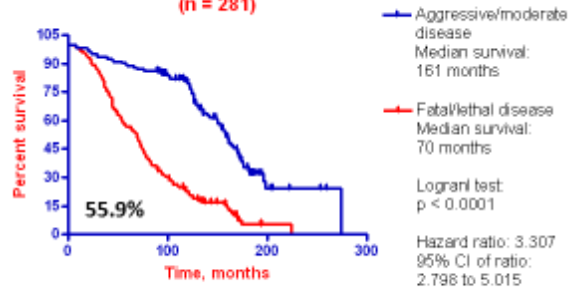

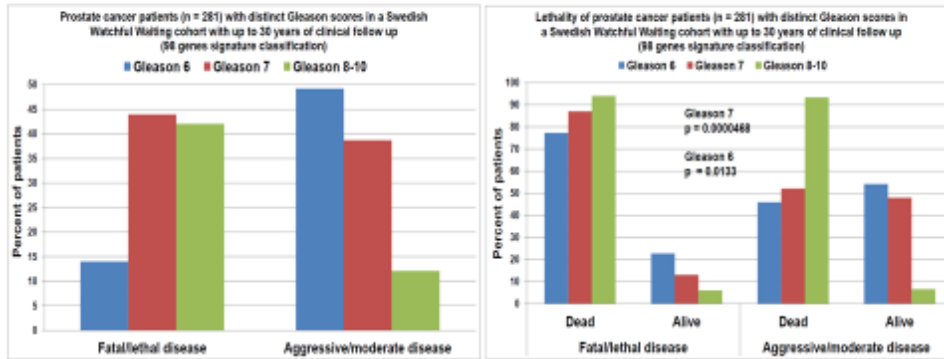

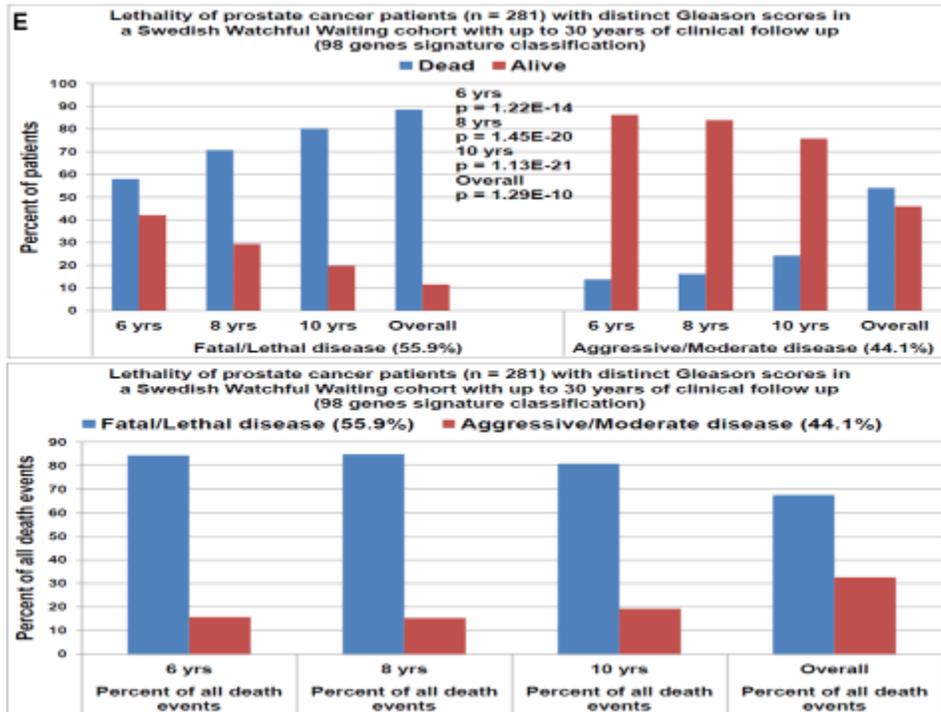

Figure 1. Natural history of prostate cancer progression in patients' population from a Swedish watchful waiting cohort with up to 30 years follow-up and classification performance of the 98 genes signature of lethal disease in prostate cancer patients.

**Figure 2.**

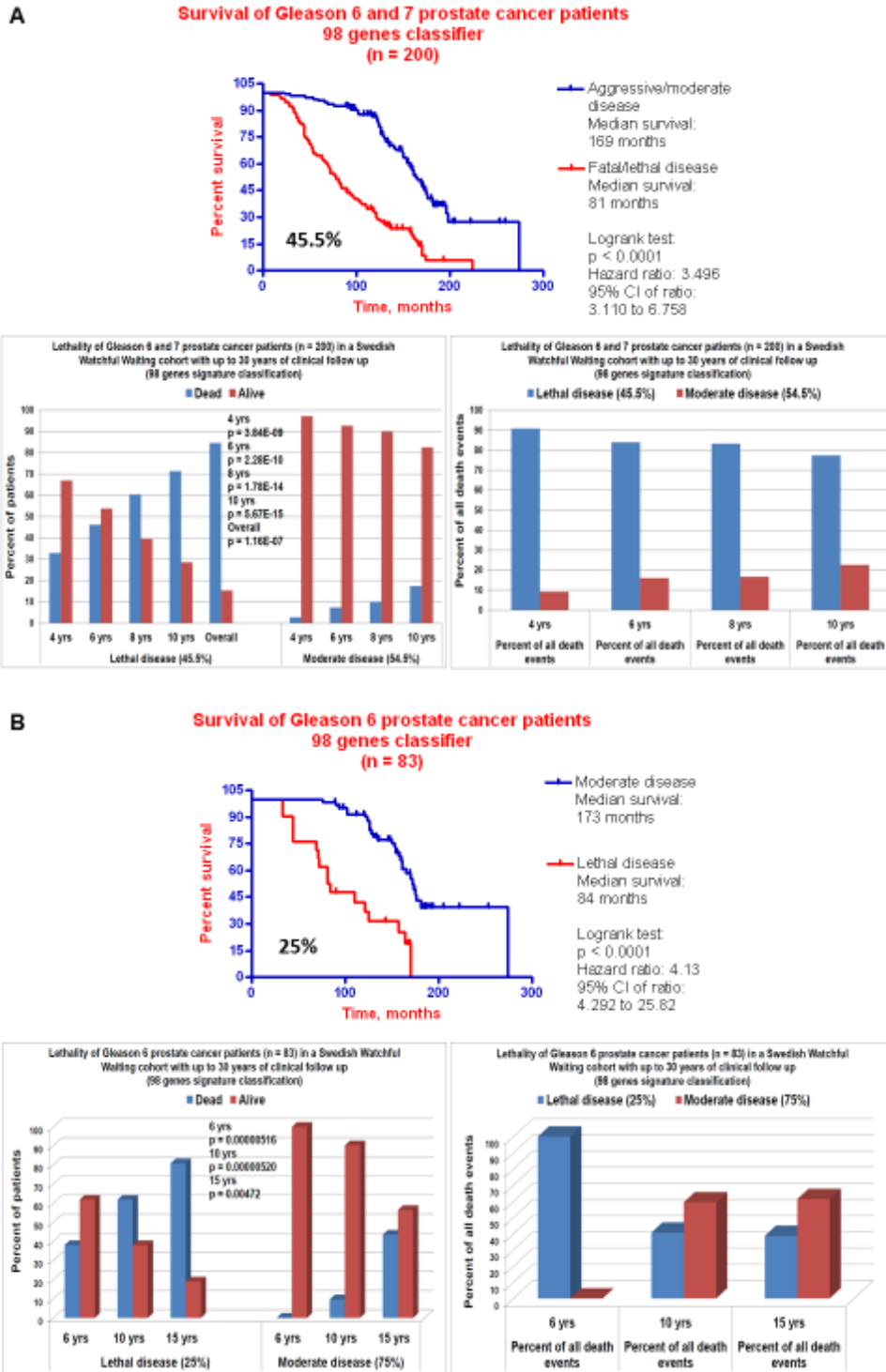

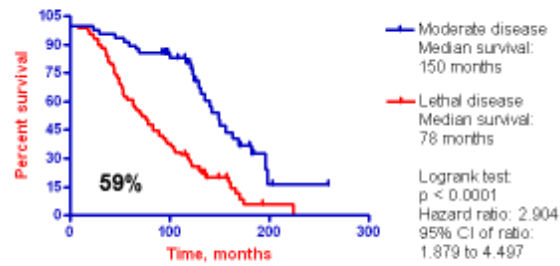
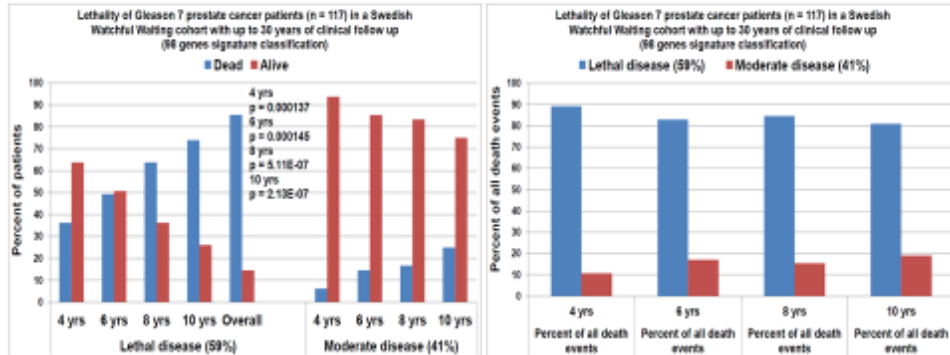

**Figure 2. Gene expression signature-based identification of lethal disease in Gleason 6 and 7 prostate cancer patients.**

**Figure 3.**

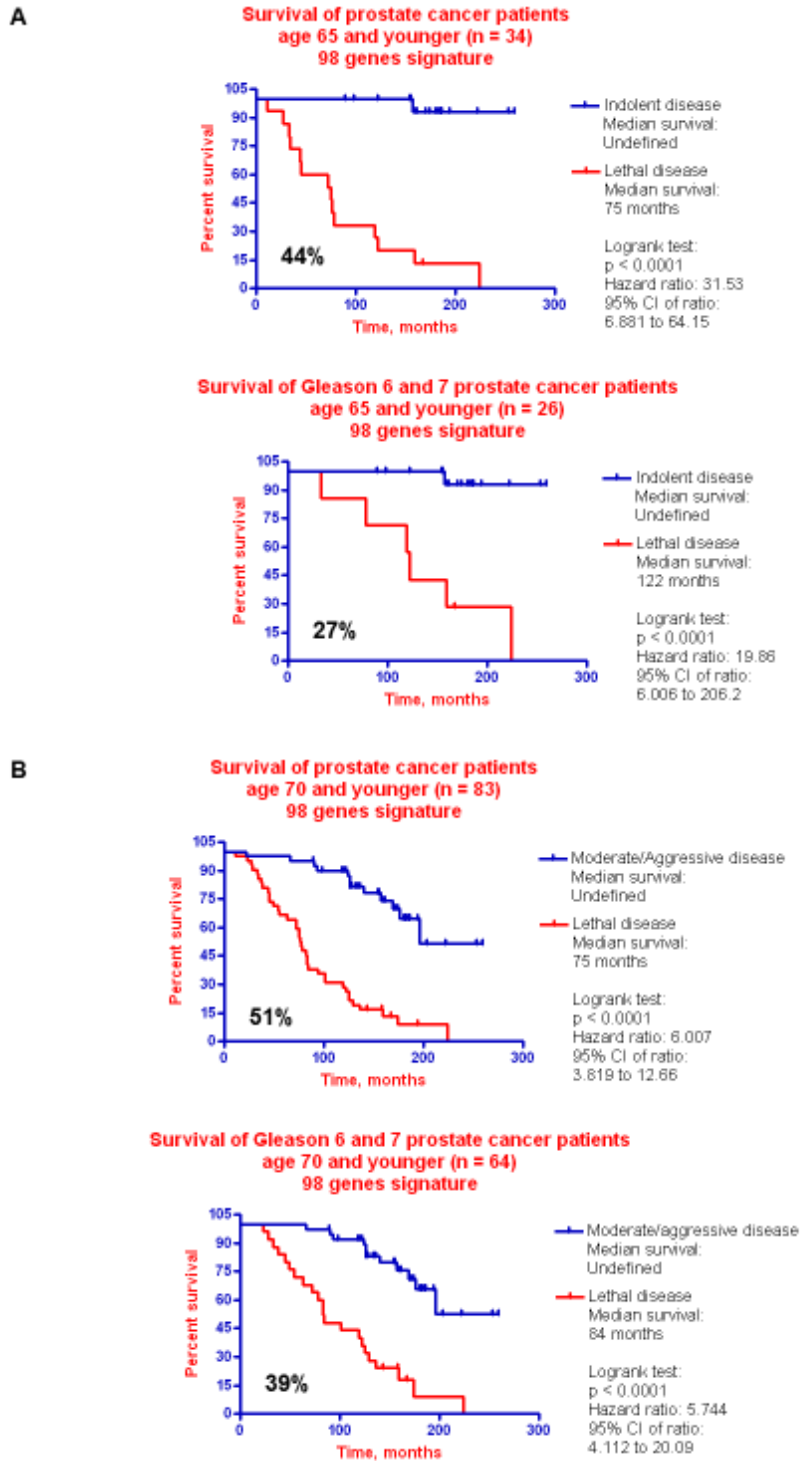

**Figure 3. Gene expression signature-based identification of lethal disease in prostate cancer patients with different age at diagnosis.**

**Figure 4.**

A
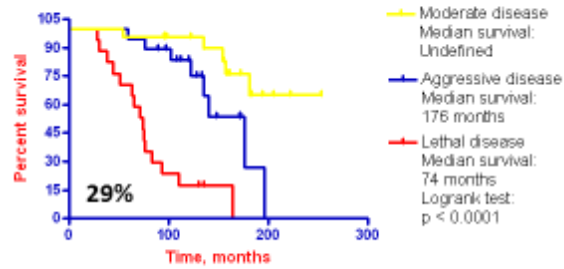
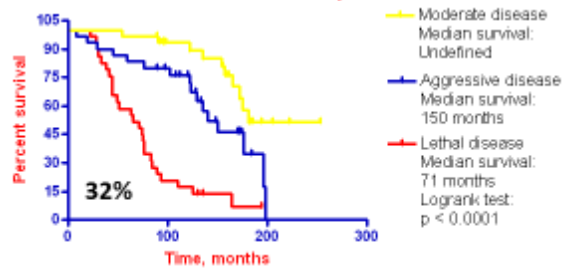

B
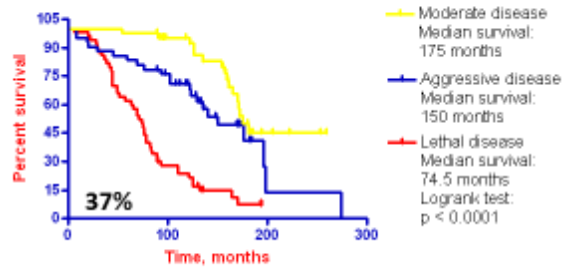
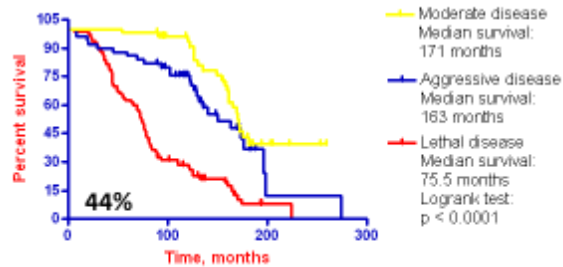

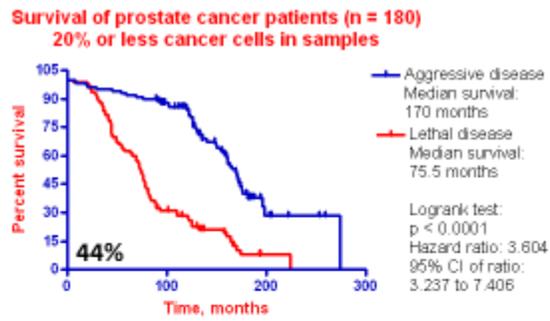

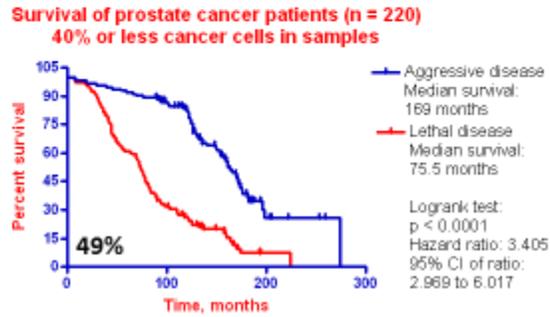

**Figure 4. Gene expression signature-based identification of lethal disease in prostate cancer patients with distinct numbers of cancer cells in biopsy samples.**

**Figure 5.**

A

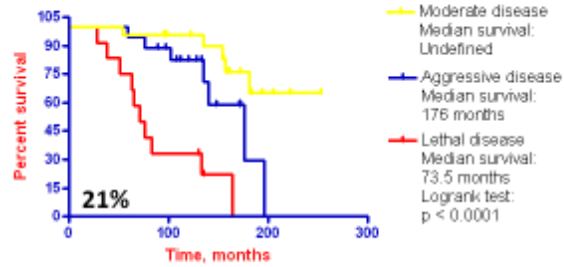

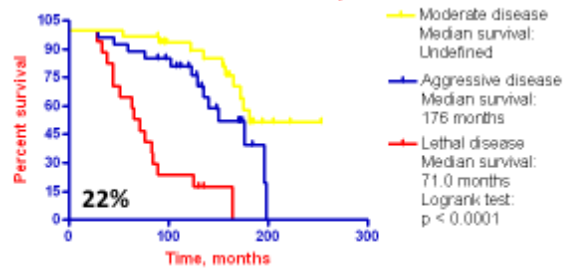

B

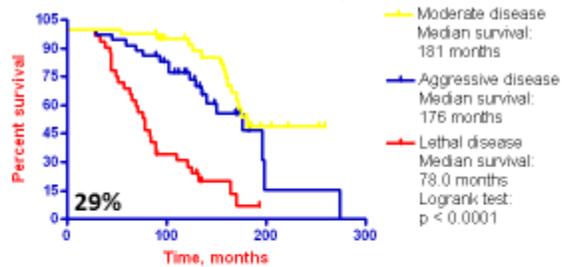

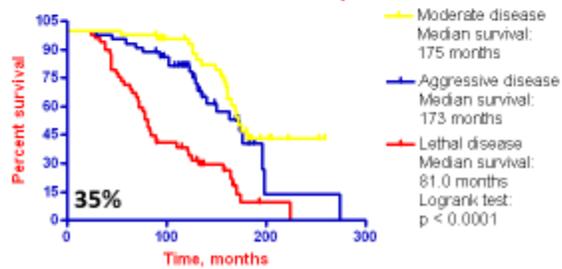

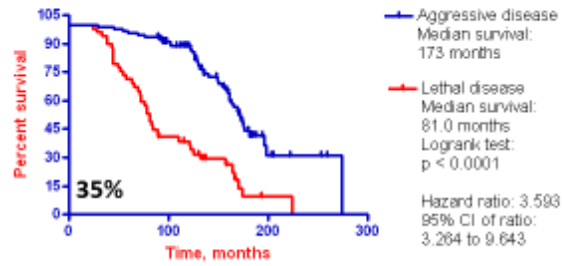

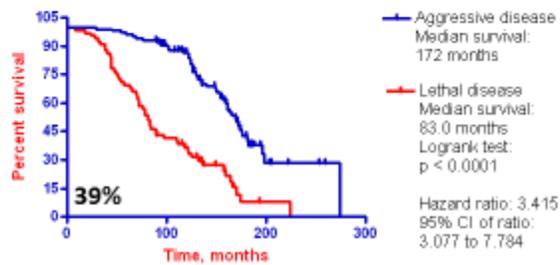

**Figure 5. Gene expression signature-based identification of lethal disease in Gleason 6 and 7 prostate cancer patients with distinct numbers of cancer cells in biopsy samples.**

**Figure 6.**

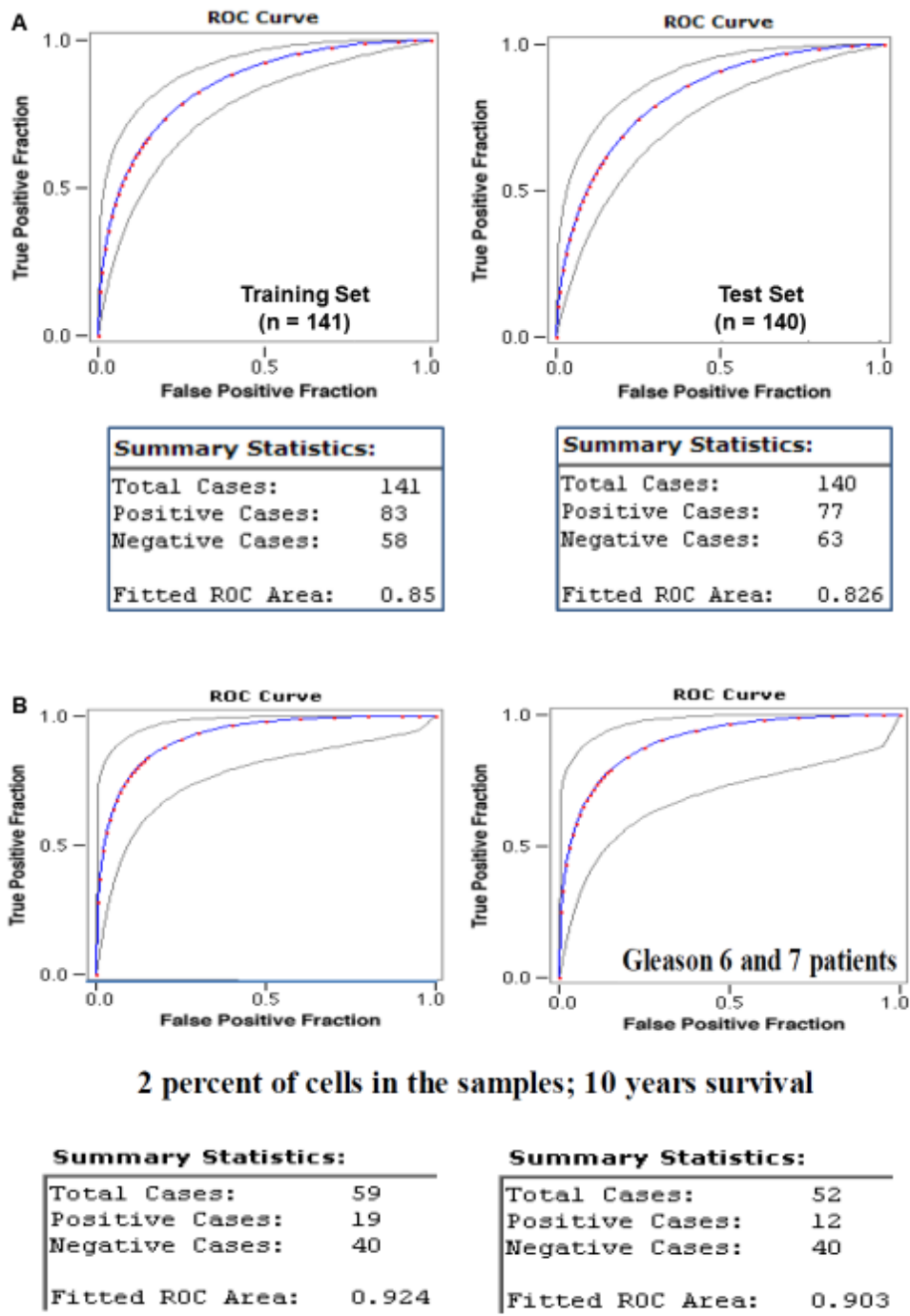

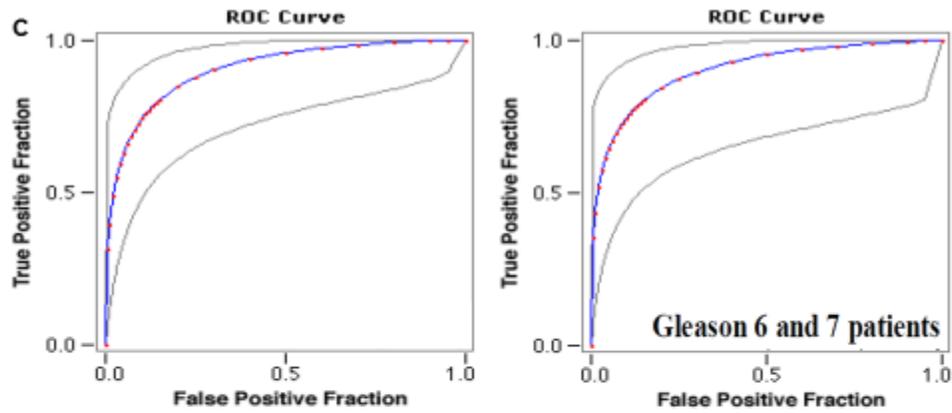

## 2 percent of cells in the samples; 6 years survival

**Summary Statistics:**

| Total Cases: | 59 |
|---|---|
| Positive Cases: | 14 |
| Negative Cases: | 45 |
| Fitted ROC Area: | 0.909 |

**Summary Statistics:**

| Total Cases: | 52 |
|---|---|
| Positive Cases: | 10 |
| Negative Cases: | 42 |
| Fitted ROC Area: | 0.906 |

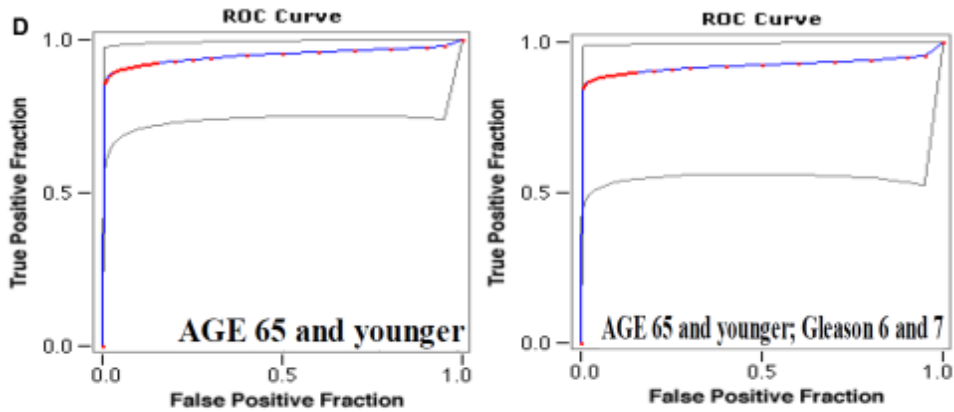

**Summary Statistics:**

| Total Cases: | 34 |
|---|---|
| Positive Cases: | 15 |
| Negative Cases: | 19 |
| Fitted ROC Area: | 0.95 |

**Summary Statistics:**

| Total Cases: | 26 |
|---|---|
| Positive Cases: | 7 |
| Negative Cases: | 19 |
| Fitted ROC Area: | 0.923 |

**Figure 6 (above).** Receiver Operating Characteristic (ROC) area under the curve analysis of the patients' classification based on the 98-genes signature score in training (n = 141) and test (n = 140) groups (A) and different clinically-relevant sub-groups (B - D) of patients.

## SUPPLEMENTAL TABLES

**Supplemental Table 1.** Univariate Cox regression analysis.

**Supplemental Table 2.** Multivariate Cox regression analysis.

**Supplemental Table 3.** Potential clinical utility of the gene expression signatures in management of active surveillance programs of prostate cancer patients with Gleason 6 and 7 tumors.

**Supplemental Table 1.** Univariate Cox regression analysis.

| | |
|---|---|
| **43 genes** | Chi Square= 11.3649; df=1; p= 0.0007 |
| **38 genes** | Chi Square= 11.2901; df=1; p= 0.0008 |
| **41 genes** | Chi Square= 11.2790; df=1; p= 0.0008 |
| **40 genes** | Chi Square= 11.0182; df=1; p= 0.0009 |
| **36 genes** | Chi Square= 10.9231; df=1; p= 0.0009 |
| 59 genes | Chi Square= 9.8677; df=1; p= 0.0017 |
| 24 genes | Chi Square= 9.8116; df=1; p= 0.0017 |
| 19 genes | Chi Square= 9.7022; df=1; p= 0.0018 |
| **22 genes** | Chi Square= 8.8170; df=1; p= 0.0030 |
| **98 genes** | Chi Square= 8.3266; df=1; p= 0.0039 |
| 35 genes | Chi Square= 7.7065; df=1; p= 0.0055 |
| Gleason | Chi Square= 6.4196; df=1; p= 0.0113 |
| 121 genes | Chi Square= 6.1059; df=1; p= 0.0135 |
| 151 genes | Chi Square= 5.5270; df=1; p= 0.0187 |
| Age | Chi Square= 4.0107; df=1; p= 0.0452 |
| 6144 genes | Chi Square= 3.1209; df=1; p= 0.0773 |

Coefficients, Std Errs, Signif, and Conf Intervs…

| | Var | Coeff. | StdErr | p | Lo95% | Hi95% |
|---|---|---|---|---|---|---|
| **43 genes** | 1 | -0.0166 | 0.0050 | 0.0008 | -0.0263 | -0.0069 |
| **38 genes** | 1 | -0.0170 | 0.0051 | 0.0009 | -0.0269 | -0.0070 |
| **41 genes** | 1 | -0.0172 | 0.0052 | 0.0009 | -0.0273 | -0.0071 |
| **40 genes** | 1 | -0.0171 | 0.0052 | 0.0010 | -0.0274 | -0.0069 |
| **36 genes** | 1 | -0.0171 | 0.0052 | 0.0010 | -0.0273 | -0.0069 |
| 59 genes | 1 | -0.0121 | 0.0039 | 0.0019 | -0.0198 | -0.0045 |
| 24 genes | 1 | -0.0203 | 0.0065 | 0.0019 | -0.0331 | -0.0075 |
| 19 genes | 1 | -0.0235 | 0.0076 | 0.0020 | -0.0384 | -0.0086 |
| **22 genes** | 1 | -0.0254 | 0.0086 | 0.0032 | -0.0422 | -0.0085 |
| **98 genes** | 1 | -0.0083 | 0.0029 | 0.0039 | -0.0140 | -0.0027 |
| 35 genes | 1 | -0.0142 | 0.0052 | 0.0059 | -0.0244 | -0.0041 |
| Gleason | 1 | -0.1381 | 0.0554 | 0.0127 | -0.2467 | -0.0295 |
| 121 genes | 1 | -0.0048 | 0.0019 | 0.0140 | -0.0086 | -0.0010 |
| 151 genes | 1 | -0.0034 | 0.0015 | 0.0193 | -0.0063 | -0.0006 |
| Age | 1 | -0.0174 | 0.0086 | 0.0443 | -0.0343 | -0.0004 |
| 6144 genes | 1 | -0.0002 | 0.0001 | 0.0737 | -0.0004 | 0.0000 |

In bold GES that outperformed clinical models in multivariate Cox regression analysis (Supplemental Table 2).

**Supplemental Table 2.** Multivariate Cox regression analysis.

Clinical model

| | | | | | |
|---|---|---|---|---|---|
| 2 co-variates model | Chi Square= 10.5052; df=2; p= 0.0052 | | | | |
| Gleason | 1 | -0.1386 | 0.0553 | 0.0122 | -0.2470 - 0.0302 |
| Age | 2 | -0.0176 | 0.0087 | 0.0425 | -0.0347 - 0.0006 |

GES models

| | | | | | |
|---|---|---|---|---|---|
| 3 co-variates model | Chi Square= 17.5914; df=3; p= 0.0005 | | | | |
| | Coefficients, Std Errs, Signif, and Conf Intervs… | | | | |
| | Var | Coeff. | StdErr | p | Lo95%    Hi95% |
| 43 genes | 1 | -0.0141 | 0.0053 | 0.0081 | -0.0245 - 0.0037 |
| Gleason | 2 | -0.0769 | 0.0595 | 0.1962 | -0.1934   0.0397 |
| Age | 3 | -0.0186 | 0.0088 | 0.0350 | -0.0360 - 0.0013 |
| 2 co-variates model | Chi Square= 15.8946; df=2; p= 0.0004 | | | | |
| 43 genes | 1 | -0.0168 | 0.0049 | 0.0006 | -0.0264 - 0.0072 |
| Age | 2 | -0.0189 | 0.0088 | 0.0327 | -0.0362 - 0.0016 |
| 2 co-variates model | Chi Square= 13.1774; df=2; p= 0.0014 | | | | |
| 43 genes | 1 | -0.0138 | 0.0054 | 0.0099 | -0.0243 - 0.0033 |
| Gleason | 2 | -0.0793 | 0.0594 | 0.1818 | -0.1956   0.0371 |
| 3 co-variates model | Chi Square= 17.3881; df=3; p= 0.0006 | | | | |
| 41 genes | 1 | -0.0145 | 0.0056 | 0.0092 | -0.0255 - 0.0036 |
| Gleason | 2 | -0.0748 | 0.0599 | 0.2119 | -0.1923   0.0426 |
| Age | 3 | -0.0186 | 0.0088 | 0.0351 | -0.0360 - 0.0013 |
| 2 co-variates model | Chi Square= 15.8063; df=2; p= 0.0004 | | | | |

| | | | | | | |
|---|---|---|---|---|---|---|
| 41 genes | 1 | -0.0174 | 0.0051 | 0.0006 | -0.0274 | -0.0074 |
| Age | 2 | -0.0189 | 0.0089 | 0.0327 | -0.0362 | -0.0016 |

2 co-variates model    Chi Square=  12.9783;  df=2;  p=   0.0015

| | | | | | | |
|---|---|---|---|---|---|---|
| 41 genes | 1 | -0.0142 | 0.0056 | 0.0111 | -0.0252 | -0.0032 |
| Gleason | 2 | -0.0773 | 0.0598 | 0.1959 | -0.1945 | 0.0399 |

3 co-variates model    Chi Square=  17.2305;  df=3;  p=   0.0006

| | | | | | | |
|---|---|---|---|---|---|---|
| 40 genes | 1 | -0.0145 | 0.0056 | 0.0101 | -0.0255 | -0.0035 |
| Gleason | 2 | -0.0755 | 0.0600 | 0.2080 | -0.1930 | 0.0420 |
| Age | 3 | -0.0188 | 0.0088 | 0.0339 | -0.0361 | -0.0014 |

2 co-variates model    Chi Square=  15.6213;  df=2;  p=   0.0004

| | | | | | | |
|---|---|---|---|---|---|---|
| 40 genes | 1 | -0.0175 | 0.0052 | 0.0007 | -0.0276 | -0.0074 |
| Age | 2 | -0.0190 | 0.0088 | 0.0313 | -0.0364 | -0.0017 |

2 co-variates model    Chi Square=  12.7583;  df=2;  p=   0.0017

| | | | | | | |
|---|---|---|---|---|---|---|
| 40 genes | 1 | -0.0141 | 0.0057 | 0.0126 | -0.0252 | -0.0030 |
| Gleason | 2 | -0.0783 | 0.0598 | 0.1906 | -0.1955 | 0.0390 |

3 co-variates model    Chi Square=  17.3697;  df=3;  p=   0.0006

| | | | | | | |
|---|---|---|---|---|---|---|
| 38 genes | 1 | -0.0144 | 0.0055 | 0.0092 | -0.0252 | -0.0036 |
| Gleason | 2 | -0.0735 | 0.0601 | 0.2213 | -0.1912 | 0.0443 |
| Age | 3 | -0.0187 | 0.0088 | 0.0344 | -0.0361 | -0.0014 |

2 co-variates model    Chi Square=  15.8509;  df=2;  p=   0.0004

| | | | | | | |
|---|---|---|---|---|---|---|
| 38 genes | 1 | -0.0172 | 0.0050 | 0.0006 | -0.0271 | - |

| | | | | | |
|---|---|---|---|---|---|
| Age | | | | | 0.0074 |
| | 2 | -0.0190 | 0.0089 | 0.0321 | -0.0363 -0.0016 |

2 co-variates model    Chi Square= 12.9222; df=2; p= 0.0016

| Var | Coeff. | StdErr | p | Lo95% | Hi95% |
|---|---|---|---|---|---|
| 38 genes | 1 | -0.0141 | 0.0056 | 0.0114 | -0.0250 -0.0032 |
| Gleason | 2 | -0.0760 | 0.0600 | 0.2050 | -0.1935 0.0415 |

3 co-variates model    Chi Square= 17.0730; df=3; p= 0.0007

| | 1 | -0.0145 | 0.0057 | 0.0108 | -0.0256 - |
|---|---|---|---|---|---|
| 36 genes | | | | | 0.0033 |
| | 2 | -0.0736 | 0.0603 | 0.2219 | -0.1918 - |
| Gleason | | | | | 0.0445 |
| | 3 | -0.0188 | 0.0088 | 0.0332 | -0.0362 - |
| Age | | | | | 0.0015 |

2 co-variates model    Chi Square= 15.5583; df=2; p= 0.0004

| | 1 | -0.0174 | 0.0052 | 0.0007 | -0.0275 - |
|---|---|---|---|---|---|
| 36 genes | | | | | 0.0073 |
| | 2 | -0.0191 | 0.0088 | 0.0307 | -0.0365 - |
| Age | | | | | 0.0018 |

2 co-variates model    Chi Square= 12.5661; df=2; p= 0.0019

| | 1 | -0.0141 | 0.0057 | 0.0138 | -0.0253 - |
|---|---|---|---|---|---|
| 36 genes | | | | | 0.0029 |
| | 2 | -0.0765 | 0.0602 | 0.2035 | -0.1945 - |
| Gleason | | | | | 0.0414 |

3 co-variates model    Chi Square= 14.8255; df=3; p= 0.0020

Coefficients, Std Errs, Signif, and Conf Intervs...

| Var | Coeff. | StdErr | p | Lo95% | Hi95% |
|---|---|---|---|---|---|
| | 1 | -0.0064 | 0.0031 | 0.0379 | -0.0125 - |
| 98 genes | | | | | 0.0004 |
| | 2 | -0.0918 | 0.0593 | 0.1219 | -0.2081 |
| Gleason | | | | | 0.0245 |
| | 3 | -0.0177 | 0.0088 | 0.0432 | -0.0349 - |
| Age | | | | | 0.0005 |

2 co-variates model    Chi Square= 12.3870; df=2; p= 0.0020

| | | Var | Coeff. | StdErr | p | Lo95% | Hi95% |
|---|---|---|---|---|---|---|---|
| 98 genes | | 1 | -0.0083 | 0.0029 | 0.0037 | -0.0139 | -0.0027 |
| Age | | 2 | -0.0177 | 0.0087 | 0.0430 | -0.0349 | -0.0006 |
| 2 co-variates model | | Chi Square= 10.7698; df=2; p= 0.0046 | | | | | |
| 98 genes | | 1 | -0.0065 | 0.0031 | 0.0374 | -0.0126 | -0.0004 |
| Gleason | | 2 | -0.0918 | 0.0593 | 0.1216 | -0.2081 | 0.0244 |
| 3 co-variates model | | Chi Square= 16.1992; df=3; p= 0.0010 | | | | | |
| | | Coefficients, Std Errs, Signif, and Conf Intervs... | | | | | |
| | | Var | Coeff. | StdErr | p | Lo95% | Hi95% |
| 22 genes | | 1 | -0.0219 | 0.0092 | 0.0174 | -0.0400 | -0.0039 |
| Gleason | | 2 | -0.0855 | 0.0593 | 0.1493 | -0.2016 | 0.0307 |
| Age | | 3 | -0.0198 | 0.0088 | 0.0251 | -0.0371 | -0.0025 |
| 2 co-variates model | | Chi Square= 14.0843; df=2; p= 0.0009 | | | | | |
| 22 genes | | 1 | -0.0270 | 0.0085 | 0.0016 | -0.0437 | -0.0103 |
| Age | | 2 | -0.0203 | 0.0088 | 0.0213 | -0.0377 | -0.0030 |
| 2 co-variates model | | Chi Square= 11.2099; df=2; p= 0.0037 | | | | | |
| 22 genes | | 1 | -0.0201 | 0.0092 | 0.0294 | -0.0382 | -0.0020 |
| Gleason | | 2 | -0.0907 | 0.0592 | 0.1253 | -0.2066 | 0.0253 |

**Supplemental Table 3.** Potential clinical utility of the gene expression signatures in management of active surveillance programs of prostate cancer patients with Gleason 6 and 7 tumors.

| Gene expression signature | Gleason sum score of eligible patients | Expected percent of patients' population | Potential clinical utility in management of active surveillance programs |
|---|---|---|---|
| 45 genes (G7) | Gleason sum 7 | 18% | Identification of patients with high likelihood of clinically fatal disease (median survival 44 months; 19% survival after 5 yrs; 100% fatality at 10 yrs) |
| 121 genes (G7) | Gleason sum 7 | 35% | Identification of patients with high likelihood of clinically lethal disease (median survival 67 months; 39% survival after 6 yrs; 90% fatality at 15 yrs) |
| 16 genes (G7) | Gleason sum 7 | 29% | Identification of patients with high likelihood of clinically lethal disease (median survival 77 months; 56% survival after 5 yrs; 94% fatality at 15 yrs) |
| 18 genes (G7) | Gleason sum 7 | 50% | Identification of patients with high likelihood of clinically lethal disease (median survival 76 months; 49% survival after 6 yrs; 22% survival after 10 yrs; 93% fatality at 15 yrs) |
| 58 genes (G6) | Gleason sum 6 | 31% | Identification of patients with high likelihood of clinically aggressive disease (median survival 150 months; 56% survival after 10 yrs; 84% fatality at 15 yrs) |
| 21 genes (G6) | Gleason sum 6 | 18% | Identification of patients with high likelihood of clinically indolent disease (93% survival after 10 yrs; 13.3% cumulative fatality) |
| 18 genes (G6) | Gleason sum 6 | 63% | Identification of patients with high likelihood of clinically aggressive disease (median survival 159 months; 94% survival after 5 yrs; 33% fatality after 10 yrs; 36% survival after 15 yrs) |
| 121 genes (G8) | Gleason sum 8-10 | 57% | Identification of patients with high likelihood of clinically fatal disease (median survival 44 months; 77% fatality after 5 yrs; 11% survival after 6 yrs; 100% fatality at 13 yrs) |

**Legend:** Gene expression signatures were developed based on a publicly available microarray analysis of a Swedish Watchful Waiting cohort with up to 30 years of clinical follow up using a novel method for gene expression profiling [cDNA-mediated annealing, selection, ligation, and extension (DASL) method] which enabled the use of formalin-fixed paraffin-embedded transurethral resection of prostate (TURP) samples taken at the time of the initial diagnosis. Details of the experimental procedure can be found in a recent publication (Sboner A, Demichelis F, Calza S, Pawitan Y, Setlur SR, Hoshida Y, Perner S, Adami HO, Fall K, Mucci LA, Kantoff PW, Stampfer M, Andersson SO, Varenhorst E, Johansson JE, Gerstein MB, Golub TR, Rubin MA, Andrén O. Molecular sampling of prostate cancer: a dilemma for predicting disease progression. BMC Med Genomics. 2010 3:8. PMID: 20233430; PMCID: PMC2855514) and in Gene Expression Omnibus (GEO: http://www.ncbi.nlm.nih.gov/geo/ ) with platform accession number: GPL5474. Full data set and associated clinical information is available at GEO with accession number: GSE16560.